\documentclass{ws-p10x7}

\begin{document}

\title{RECENT PROGRESS IN HEAVY QUARK PHYSICS}

\author{Mark B. Wise}

\address{Theoretical Physics, Caltech, Pasadena, CA 91125,
USA\\E-mail: wise@theory.caltech.ed}

\twocolumn[\maketitle\abstract{Some of the recent progress in heavy quark physics is reviewed. Special attention is paid to inclusive methods for determining $V_{ub}$ and factorization in nonleptonic $\bar B$ decays. Theoretical predictions for $\bar t t$ production near threshold are also discussed.    }]

\section{Model Independent Predictions}

Since QCD cannot be solved exactly every theoretical prediction of the properties of hadrons containing one or more heavy quarks involves approximations of some kind. Much of the theoretical progress in the last ten years has arisen from treating the heavy quark mass $m_Q$ as large and expanding in $\Lambda_{QCD}/m_Q$ and $\alpha_s(m_Q)$. In this talk $\Lambda_{QCD}$ denotes a nonperturbative quantity of order the scale at which $QCD$ becomes strongly coupled. Some such quantities are:
$f_{\pi} \sim 140 ~{\rm MeV}$, $m_{\rho} \sim 770 ~{\rm MeV}$ and $m_K^2/m_s \sim 2 ~{\rm GeV}$. Clearly with $m_c \sim 1.5~{\rm GeV}$ and $m_b \sim 4.8~{\rm GeV}$ we cannot have complete confidence in the first few terms of such expansions. Experimental guidance is needed to see in which cases they work. Notice that 
$m_K^2/m_s$ is quite large. Sometimes expressions containing this factor are said to be chirally enhanced. But this is misleading. The ratio $m_K^2/m_s$ is finite as $m_s \rightarrow 0$ and is not enhanced parametrically compared with the other two quantities. It is just of order $\Lambda_{QCD}$. In fact if you were to ask a very bright theorist who knows nothing about experimental data\footnote{It is not hard to find such an individual. Almost any young string theorist will do.} which of these three quantities is large he (or she) would pick
$f_{\pi}$ because it goes to infinity in the large number of colors limit while the other two stay fixed.

A similar situation holds for lattice QCD. There one works at finite spatial volume with a finite lattice spacing. To treat the heavy quarks dynamically ({\it i.e.} no $\Lambda_{QCD}/m_Q$
expansion) we need $1/a \gg m_Q$, where $a$ is the lattice spacing. To take into account the long distance nonperturbative effects  one needs $L \gg 1/\Lambda_{QCD}$, where $L$ is the length of a spatial dimension. Combining these, the number of lattice sites in one dimension $N=L/a$ must be much larger than $\xi_Q=m_Q/\Lambda_{QCD}$. Numerically, with $\Lambda_{QCD}$ set equal to $200~ {\rm MeV}$,  $\xi_c \sim 8$ and $\xi_b \sim 24$. Lattice QCD predictions for heavy quark systems use an expansion in $\xi_Q/N$.

It is also important to remember that many predictions also rely on "lore". Approximations we believe are valid at some level but for which we don't understand how to quantify the corrections to. For example, "local" duality, which is used for making predictions for inclusive $\bar B$ and $\Lambda_b$ decays ({\it e.g.} lifetimes). The quenched approximation in lattice QCD also falls into this category.

In this talk I restrict the references that I give to papers that appeared in the year 2000 or later. Even with this strong cut the number of papers cited will exceed $20$.

\section{$|V_{ub}|$ from Inclusive $\bar B$ Semileptonic Decay}

Predictions for inclusive $\bar B \rightarrow X e \bar \nu_e$ differential decay rates are made using the operator product expansion (OPE) and the $\Lambda_{QCD}/m_Q$ expansion.
The fully differential decay rate is $d^3\Gamma/dE_edE_{\bar \nu_e}dq^2$, where $q=p_e+p_{\bar \nu_e}$. Usually one considers single variable distributions formed by integrating this over two of its variables ({\it e.g.} the electron energy distribution $d\Gamma/dE_e$). Even after integrating over two of the kinematic variables not all regions of phase space can be predicted using the first few terms in the operator product and $\Lambda_{QCD}/m_Q$ expansions. Intuition about when one runs into trouble by focusing on the first few terms of the OPE can be gained from some simple kinematic considerations. 

At fixed $m_X$ the minimum value of $q^2$ occurs when the electron and neutrino are parallel and $X$ recoils against them. The maximum value of $q^2$ is $(m_B-m_X)^2$ and it occurs for the configuration where the state $X$ is at rest and the electron and its anti-neutrino are back to back. Removing the fixed $m_X$ constraint the region 
\begin{equation}
(m_B-m_D)^2 \le q^2 \le m_B^2,
\end{equation}
must come from the $b \rightarrow u$ transition.

At fixed $m_X$ the minimum value of the electron energy $E_e$ is zero while the maximum value is $(m_B^2-m_X^2)/(2m_B)$. The maximum value of the electron energy occurs (for $m_X \ne 0$) when the neutrino carries no energy and the electron and the hadron $X$ are back to back. Removing the fixed $m_X$ constraint the region
\begin{equation}
(m_B^2-m_D^2)/(2m_B) \le E_e \le m_B/2,
\end{equation}
must come from the $b \rightarrow u$ transition.

Physically it is clear that for a rapidly recoiling states $X$ differential decay rates that are dominated by the region of final hadronic masses $\Delta m_X^2 \sim \Lambda_{QCD}m_B$ are very sensitive to nonperturbative QCD and will not be calculable using a few terms in the OPE. On the other hand for final hadronic states that are almost at rest this sensitivity to nonperturbative physics occurs for the region $\Delta m_X \sim \Lambda_{QCD}$. These regions of hadronic invariant mass correspond in the first case to $E_e$ within $\Lambda_{QCD}$ of the endpoint and in the second case $q^2$ within $m_B \Lambda_{QCD}$ of the endpoint.

\begin{figure}
\epsfxsize20pc
\figurebox{}{pc}{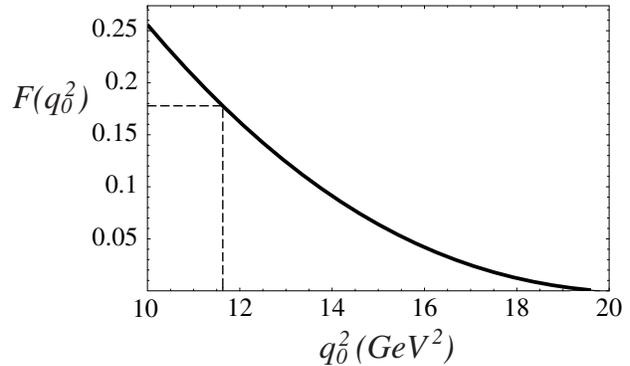}
\caption{The fraction of $b \rightarrow u$ events in the region $q^2\ge q_0^2$ plotted as a function of $q_0^2$}
\label{fract} 
\end{figure}

Bauer, Ligeti and Luke noted\cite{bll1,bn} that this is good news for using a measurement of the $q^2$ spectrum to get $|V_{ub}|$.  Since $2m_B m_D \gg m_B \Lambda_{QCD}$ a prediction for, $F(q_0^2)$, the fraction of $b \rightarrow u$ events expected in the region $q^2\ge q_0^2$, can be made using the first few terms in the OPE. The $b$ to $u$ mixing angle can be extracted from a measurement of the branching fraction of events in that region, ${\cal B}(\bar B\to X_u e\bar\nu_e)|_{q^2 > q_0^2}$, using
\begin{equation}
|V_{ub}| = 3.04\times 10^{-3}\,
  \left( {{\cal B}(\bar B\to X_u e\bar\nu_e)|_{q^2 > q_0^2} \over 
  0.001 \times F(q_0^2) }\right)^{1 \over 2}.
\end{equation}
Figure (\ref{fract}) shows a prediction for $F(q_0^2)$ plotted as a function of the cut $q_0^2$.
The dashed line indicates the cut $q_0^2=(m_B-m_D)^2 = 11.6={\rm GeV^2}$, which corresponds to $F=0.178$. Perturbative effects of order $\alpha_s^2\beta_0$ and the nonperturbative effects characterized by $\bar \Lambda$, $\lambda_1$ and $\lambda_2$ have been included in this prediction for $F(q_0^2)$. These nonperturbative parameters  occur in any prediction for  inclusive $\bar B$ decay. The values, $\bar \Lambda =0.35 \pm 0.13$ and $\lambda_1=-0.238 \pm 0.11$ have recently been extracted by the CLEO collaboration\cite{cleo1} from experimental data on the hadronic mass distribution in $\bar B\to X_c e\bar\nu_e$ and the photon energy spectrum in weak radiative $\bar B$ decay. It may be possible to reduce the theoretical uncertainties and increase the amount of phase space considered by combining cuts\cite{bll2} in $q^2$ and $m_X^2$.

Even though effects of dimension five operators have been included in the prediction of $F(q_0^2)$ there is still a significant uncertainty from even higher dimension operators. Voloshin has recently stressed the possible importance of the dimension six four quark operators\cite{voloshin}. 
\begin{figure}
\epsfxsize15pc
\figurebox{}{pc}{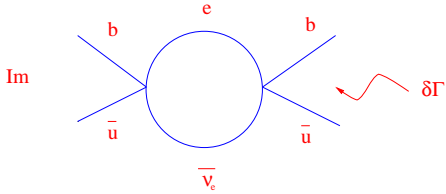}
\caption{Feynman diagram used to determine the contribution of four quark operators to the OPE for the semileptonic decay rate. Im denotes that it is the imaginary part that is relevant.}
\label{voloshin}
\end{figure}
A calculation of the imaginary part of the Feynman diagram in Figure (\ref{voloshin}) reveals that they contribute
\begin{equation}
\left(\delta \Gamma \over \Gamma_{SL} \right)(\bar B \rightarrow X_u e \bar \nu_e) \simeq \left(4 \pi f_B \over m_B \right)^2~(B_1-B_2)\,
\end{equation}
to the semileptonic rate. In the above equation the dimensionless parameters $B_{1,2}$ determine the matrix elements of the two relevant four-quark operators. Explicitly,
\begin{equation}
\langle \bar B|(\bar b_L \gamma^{\mu} u_L)(\bar u_L \gamma_{\mu} b_L)|\bar B \rangle = B_1~{f_B^2 m_B \over 8},
\end{equation}
and 
\begin{equation}
\langle \bar B|(\bar b_R u_L)(\bar u_L b_R)|\bar B \rangle = B_2~{f_B^2 m_B \over 8}.
\end{equation}

Note that in the vacuum insertion approximation $B_1=B_2=1$ and the effect of these four quark operators vanishes. It is easy to understand why this is the case. In the vacuum insertion approximation the b quark and anti-up quark on the left side of Figure (\ref{voloshin}) must be in the initial state $\bar B$ meson. So the calculation reduces to the  square of the amplitude for $\bar B \rightarrow e \bar \nu_e$ which vanishes when the electron mass is neglected. 

Numerically
\begin{equation}
 \left(\delta \Gamma \over \Gamma_{SL} \right)\simeq 0.02 \left(f_B \over 0.2 {\rm GeV} \right)^2~\left({B_1-B_2 \over 0.1}\right),
\end{equation}
so if $B_1-B_2 \sim 0.1$ these operators contribute about $2\%$ to the semileptonic rate.  But note that this contribution corresponds to the kinematic situation where the electron and anti-neutrino have energies about half the $B$ mass and the final hadronic state has low mass and low energy of order $\Lambda_{QCD}$. Consequently their contribution is concentrated in the region $q^2 \ge (m_B-m_D)^2$. If $20\%$ of the $q^2$ spectrum comes from this region then, for $B_1-B_2 \sim 0.1$ , these operators cause at least a $5\%$ uncertainty in the extraction of $|V_{ub}|$ from the $q^2$ spectrum. It is possible we will learn about the importance of these four quark operators from experimental data. For example, they give different contributions to charged and neutral $B$ semileptonic decay.

The endpoint region of the electron energy spectrum also comes from  the $b \rightarrow u$ transition. However the rate in this region is probably not dominated by the first few operators in the OPE. The effects of operators that dominate in a region $\Delta E_e \sim \Lambda_{QCD}$ of the endpoint can be summed into a shape function. Neglecting perturbative QCD effects and effects suppressed by $\Lambda_{QCD}/m_b$,
\begin{equation}
{d \Gamma_{SL} \over dx_b}={G_F^2 |V_{ub}|^2 m_b^5 \over 96 \pi^3}S_{SL}(x_b)
\end{equation}
where $x_b=2E_e/m_b$ and
\begin{equation}
S_{SL}(y)=\langle \bar B(v)|\bar b_v\theta(1-y+in \cdot D/m_b)b_v|\bar B(v)\rangle
\end{equation}
with $n$ a light-like four-vector and $D$ a covariant derivative. So to determine $|V_{ub}|$ from the endpoint of the electron energy spectrum one needs to know the integral of $S_{SL}$ over the endpoint region. Fortunately the required integral can be determined from the photon energy spectrum in weak radiative $\bar B$ decay. Again neglecting perturbative effects and terms suppressed by $\Lambda_{QCD}/m_b$,
\begin{equation}
{d \Gamma_{WR} \over dx_b}={G_F^2 \alpha |C_7^{(0)}|^2|V_{ts}^*V_{tb}|^2 m_b^5 \over 32 \pi^4}S_{WR}(x_b)
\end{equation}
where $C_7^{(0)}=-0.31$ is the leading order Wilson coefficient of the transition magnetic moment operator in the effective Hamiltonian for weak radiative decay and $S_{WR}$ is the shape function appropriate to weak radiative decay,
\begin{equation}
S_{WR}(y)=\langle \bar B(v)|\bar b_v\delta(1-y+in \cdot D/m_b)b_v|\bar B(v)\rangle.
\end{equation}
The two shape functions are not equal. However, it is easy to derive a relationship between the two since
\begin{equation}
\label{derivative}
d \theta(1-y+in \cdot D/m_b)/dy=-\delta(1-y+in \cdot D/m_b).
\end{equation}
Using equation (\ref{derivative}) yields
\begin{eqnarray}
\int_x dy(y-x)S_{WR}(y) &=&-\int_x dy(y-x){d S_{SL}(y) \over dy} \nonumber  \\
&=&\int_x dyS_{SL}(y)
\end{eqnarray}
so a weighted integral of the weak radiative decay shape function is equal to the integral of the semileptonic shape function. Putting these results together and noting that, $|V_{ts}^*V_{tb}|^2\simeq |V_{cb}|^2$, gives
\begin{eqnarray}
\label{vub}
{|V_{ub}|^2 \over |V_{cb}|^2}&=&{3 \alpha |C_7^{(0)}|^2 \int_x  (d\Gamma_{SL}/dy)dy \over \pi \int_x (y-x)(d\Gamma_{WR}/dy)dy } \nonumber \\
 &+& {\cal O}(\alpha_s)+{\cal O}({\Lambda_{QCD} \over m_b})
\end{eqnarray}
It is important to include the perturbative corrections. They are singular in the endpoint region, but the most singular pieces can be resumed\cite{llr}. Including them changes the weighting function in the formula above.  At this order the other operators in the effective Hamiltonian also enter\cite{n}. All together the perturbative effects result in about a $10\%$ increase in the value of $|V_{ub}|$ over what equation (\ref{vub}) would imply.

Unknown order $\Lambda_{QCD}/m_b$ corrections naively imply about a $5\%$ theoretical uncertainty in the value of $|V_{ub}|$ extracted in this way\footnote {This estimate is not more sophisticated than the methods used by my doctor for  my back problem. You take $\Lambda_{QCD} \sim 500~{\rm MeV}$ and $m_b \sim 5 {\rm GeV}$ implying a $10\%$ uncertainty in equation (\ref{vub}) or equivalently a $5\%$ uncertainty in the value of $|V_{ub}|$.}. There are also additional uncertainties from possible violations of duality. For example, if the endpoint region of the electron spectrum is dominated by just the $\pi$ and $\rho$ final states I would be very suspicious of the use of results based on the OPE. Even within the OPE approach higher dimension operators may give effects that are larger than the naive $5\%$ estimate. For example, if $10\%$ of the semileptonic $b \rightarrow u$ events are beyond the endpoint energy cut at $2.3~{\rm GeV}$ and $B_1-B_2 \sim 0.1$ then the dimension six four quark operators induce a $10\%$ uncertainty in the determination of $|V_{ub}|$ with this method.

If the inclusive semileptonic $b \rightarrow u$ rate is measured without making any cuts that restrict you to regions of phase space where higher dimension operators are important then $|V_{ub}|$ can be determined with small theoretical uncertainty. However, it is difficult to imagine that being done with the large $b \rightarrow c$ background. The LEP groups report measurements of the totally inclusive rate. For example, the OPAL collaboration reports\cite{opal} ${\cal B}( \bar B \rightarrow X_u e \bar \nu_e)=(1.6 \pm 0.8)\times 10^{-3}$ giving $|V_{ub}|$ at the $25\%$ level. It is not completely clear to me what region of phase space is emphasized to remove the $b \rightarrow c$ background, but if it is the low hadronic invariant mass region then there is a sizeable theoretical uncertainty because the rate in this region is not given by the lowest dimension operators in the OPE.

In the future I am hopeful that $|V_{ub}|$ will be known at the $5-10\%$ level. Confidence in the precision will come from consistency between several different model independent methods used to determine it. This includes predictions for exclusive decays from Lattice QCD which are likely to also play an important role in this program. 

\section{Factorization for $\bar B \rightarrow D^{(*)}X$ }

Exclusive nonleptonic $K$ and $D$ weak decays have proven to be very difficult to understand using systematic methods. In kaon decays there is the factor of twenty enhancement of $\Delta I=1/2$ amplitudes, that has no simple explanation but rather is thought to come from several different sources of enhancement. Working on this subject has ruined many a promising career in theoretical physics. My advice to graduate students has always been: If you drink from the nonlep tonic your physics career will be ruined and you will end up face down in the gutter\footnote{I believe that the phrase "drinking the nonlep tonic" was originally introduced by H. Lipkin}. 

However, in $B$ decays it seems clear that in some situations a systematic approach to understanding nonleptonic decay amplitudes is possible. For decays of the form $\bar B \rightarrow D^{(*)} X$ where $X$ is a low mass hadronic state there is a strong theoretical argument, valid to all orders in perturbation theory\cite{bps}, that factorization is the leading term in the systematic expansion of these amplitudes in powers of $\Lambda_{QCD}/m_Q$ and $\alpha_s(m_Q)$. Furthermore the perturbative corrections are computable.

The effective Hamiltonian for Cabibbo allowed $\bar B \rightarrow D^{(*)}X$ nonleptonic decays is
\begin{equation}
{\cal H}={ 4G_F \over \sqrt 2}V_{ud}^*V_{cb}\left[C_0(\mu)O_0(\mu)+C_8(\mu)O_8(\mu)\right]
\end{equation}
where
\begin{equation}
O_0=\bar c_L\gamma^{\mu}b_L \bar d_L \gamma_{\mu}u_L
\end{equation}
and
\begin{equation}
O_8=\bar c_L\gamma^{\mu}T^Ab_L \bar d_L \gamma_{\mu}T^Au_L.
\end{equation}
The Wilson coefficients $C_0(\mu)$ and $C_8(\mu)$ are known at the next to leading logarithmic level. It is convenient to introduce,
\begin{equation}
\label{hat}
\hat C_0 = C_0+{\alpha_s(\mu) \over 18\pi}\left[-6 {\rm  ln} (\mu^2/m_b^2) +B\right]C^{(0)}_8.
\end{equation}
Here I have moved some universal corrections to matrix elements into the coefficient so that $\hat C_0(\mu)$ is independent of the subtraction point $\mu$ at the next to leading logarithmic level\footnote{It has a very small residual subtraction point dependence.}. In equation (\ref{hat}) $B$ is a scheme dependent constant. The factorization formula for the decay amplitudes we are considering is,
\begin{eqnarray}
\langle D^{(*)}X|{\cal H}|\bar B \rangle&=&{4 G_F \over \sqrt 2}V_{ud}^*V_{cb}\hat C_0(m_b) \hat M 
\nonumber  \\
&+& {\cal O}(\alpha_s) +{\cal O} ({\Lambda_{QCD} \over m_b}),
\end{eqnarray}
where
\begin{equation}
\hat M=\langle D^{(*)}|\bar c_L \gamma^{\mu} b_L|\bar B \rangle  \langle X|\bar u_L \gamma_{\mu} d_L|0 \rangle.
\end{equation}

\begin{figure}
\epsfxsize15pc
\figurebox{}{pc}{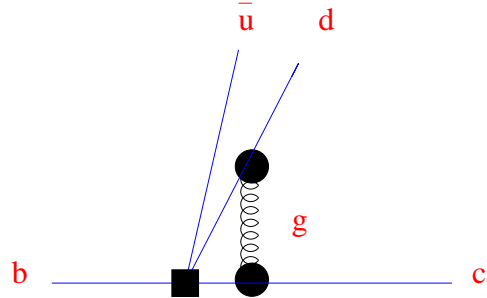}
\caption{One loop Feynman diagram that contributes to the order $\alpha_s$ correction to factorization. It has an imaginary part.}
\label{fact}
\end{figure}

The matrix element $\langle D^{(*)}|\bar c_L \gamma^{\mu} b_L|\bar B \rangle$ is measured in semileptonic $\bar B$ decay while the matrix element $\langle X|\bar u_L \gamma_{\mu} d_L|0 \rangle$ is often known from experiments involving $X$. For example, when $X$ is a pion that matrix element is characterized by the pion decay constant which determines the charged pion lifetime.

The order $\alpha_s(m_b)$ perturbative QCD corrections have been calculated\cite{bbns}. The situation here is much like the Brodsky, Lepage formalism for exclusive processes (e.g. the pion electromagnetic form factor $F_{\pi}(Q^2)$.) The order $\alpha_s(m_b)$ correction  is expressed as a hard scattering amplitude convoluted with the light cone distribution for $X$. Using $\phi_{\pi}(x)=6x(1-x)$ it corrects the $\bar B \rightarrow D \pi$ amplitude
by the factor $0.014 +0.014i$. It is small because it is suppressed by $1/N_c^2$. Notice that there is a calculable imaginary part. That is not surprising. One loop Feynman diagrams like that in Figure (\ref{fact}) where a gluon goes between the charm quark and one of the quarks that goes into the pion contribute to this correction and they clearly have an imaginary part.

One important difference between $\bar B \rightarrow D^{(*)} X$ and the pion electromagnetic form factor is that at leading order the large momentum transfer is provided by the weak Hamiltonian and not by a hard gluon. This avoids the familiar problem with the pion electromagnetic form factor. It takes quite large momentum transfers before nonperturbative effects are small enough that the expression for $F_{\pi}(Q^2)$ in terms of a hard scattering amplitude convoluted with pion light cone distributions is valid. For these large $Q^2$ $F_{\pi}(Q^2)$ is too small to be measured.

Its interesting to note that for final states $X$ with spin greater than one the leading factorized amplitude vanishes, but at order $\alpha_s(m_b)$ there will be a calculable contribution\cite{dh}. Unfortunately there will also be incalculable order $\Lambda_{QCD}/m_Q$ contributions as well. 

\begin{figure}
\epsfxsize17pc
\figurebox{}{pc}{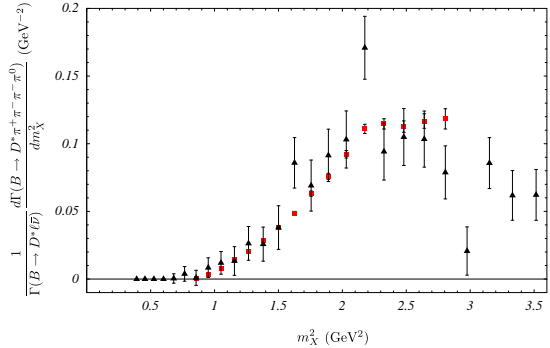}
\caption{Factorization prediction for $d\Gamma(\bar B^0 \rightarrow D^{*+}\pi^+\pi^-\pi^-\pi^0)/dm_X^2$ (the squares) compared with $B$ decay data (the triangles)}
\label{fpi}
\end{figure}

For $\bar B^0 \rightarrow D^{(*)+} X^-$, when the low mass hadronic final state $X$ is a $\pi$ or a $\rho$, factorization has been checked at the $5\%$ level (in the matrix elements). Some nonperturbative corrections are expected to grow with the mass of $X$ and should be suppressed by $m_X/E_X$. It would be interesting to detect these corrections. One approach is to use many body final states where the $X$ invariant mass can be varied\cite{llw}. The very accurate $\tau$ decay data\cite{cleo2} can be used to measure $\langle X|\bar u_L \gamma_{\mu} d_L|0 \rangle$ for multibody final states $X$ and combining this result with semileptonic $\bar B$ decay one gets the prediction of factorization for $\bar B^0 \rightarrow D^{(*)+} X^-$. This can be compared with  $\bar B^0 \rightarrow D^{(*)+} X^-$ decay data over a range of $m_X$ . Since, $m_{\tau}/E_X \sim 0.7$, is not small nonperturbative corrections may be observable over the range of $m_X$ that $\tau$ decay data can probe. A plot of the prediction of factorization (using  $\tau$ decay and $\bar B$ semileptonic decay data) along with the $\bar B^0$ decay data is shown in Figure (\ref{fpi}) for the differential decay rate $d\Gamma(\bar B^0 \rightarrow D^{*+}\pi^+\pi^-\pi^-\pi^0)/dm_X^2$. The factorization prediction is shown as squares and the $\bar B^0$ decay data is shown with triangles. At the present time there is no evidence for violations of factorization that grow as $m_X$ increases, but much higher precision data should be available in the future.

One problem with this comparison, that might arise as the $B$ decay data gets more precise, is that the $\pi^+\pi^-\pi^-\pi^0$ final state can arise not just from the weak current $\bar u_L \gamma_{\mu} d_L$. Three pions can be produced off the weak current and an excited $D$ from the $b \rightarrow c $ current which then decays to $D^* \pi$ giving the four-pion final state. One way to test the importance of this process is to measure
\begin{equation}
R_{0-}= {\Gamma (\bar B^0 \rightarrow D^{*0} \pi^+\pi^+\pi^-\pi^-)\over \Gamma (\bar B^0 \rightarrow D^{*+} \pi^+\pi^-\pi^-\pi^0)}.
\end{equation} 
The final state $\pi^+\pi^+\pi^-\pi^-$ cannot come from the weak current $\bar u_L \gamma_{\mu} d_L$, since its total charge is neutral and this current produces a final state with charge $-1$. Experimentally\cite{cleo3}, $R_{0-}= 0.17 \pm 0.04 \pm 0.02$ and  $R_{0-} \le 0.13$ at $90\%$ in the region $m_X^2 \le 2.9 GeV^2$.

Factorization predicts that
\begin{equation}
{\Gamma(B^- \rightarrow D^0 \pi^-) \over \Gamma(\bar B^0 \rightarrow D^+ \pi^-)}=1+{\cal O} ({\Lambda_{QCD} \over m_Q}).
\end{equation}
The process where the spectator quark in the $B^-$ goes into the $\pi^-$ is an order $\Lambda_{QCD} /m_Q$ correction.  A naive estimate suggests that the $\Lambda_{QCD}/m_Q$ suppression  is not effective (it is compensated by other factors like the ratio $f_D/f_{\pi}$, which is formally of order $(\Lambda_{QCD} /m_c)^{1/2}$  but is actually expected to be greater than unity). However, one might still expect this ratio to be very near unity since the amplitude where the spectator quark goes into the pion is {\it color suppressed}. Experimentally the above ratio $1.8 \pm 0.3$. 

Similarly  the amplitude for $\Gamma(\bar B^0 \rightarrow D^0 \pi^0)$ should be suppressed by $\Lambda_{QCD}/m_Q$. Unfortunately these suppressed terms cannot be computed. Two recent measurements are\cite{cleo4,belle}: ${\cal B}(\bar B^0 \rightarrow D^0 \pi^0)=[2.7\pm 0.3 \pm 0.6]\times 10^{-4}$ and ${\cal B}(\bar B^0 \rightarrow D^0 \pi^0)=[3.1 \pm 0.4 \pm 0.5] \times 10^{-4}$. 
\section{Calculating $\bar B \rightarrow M M$ amplitudes }

The ideas presented in the previous section have been extended to nonleptonic $\bar B$ decays involving two light final states. Understanding these decay amplitudes is very important for studying CP nonconservation. There are two approaches to this problem. The difference hinges on the effectiveness of Sudakov suppression of the endpoint region. It's a little like the old debate over the validity of the prediction using the formalism of Brodsky and Lepage for $F_{\pi}(Q^2)$ at the values of $Q^2$ where it is measured. One group assumes this Sudakov suppression is effective\cite{kls} and the other arrives at a different power counting since they assume it is not effective\cite{bbns}. My own personal preference would be to side with the latter. One prediction of this approach ({\it i.e.} the one where the Sudakov suppression of the endpoint region is assumed not to be effective) is that for these decays final state strong phases should be small. 

Unfortunately it is clear that some formally suppressed effects are actually very important. They are the ones that are suppressed by $m_K^2/(m_s m_Q)$ and are called chirally enhanced. However, as I remarked in the first section they are not enhanced by any parameter of QCD. Without a complete theory of the order $\Lambda_{QCD}/m_Q$ corrections it hardly seems systematic to include them in the predictions that are being made. Recently there have been suggestions in the literature that effects that are associated with Feynman diagrams involving a charm loop might not be described adequately using perturbation theory\cite{b,cfmps}.

It is too early to tell how useful this approach will be. It is certainly important to better understand the effects 
that are classified as subleading and to find other areas where it can be applied and tested. One recent example, is the exclusive weak radiative decay\cite{bfs,bb} $\bar B \rightarrow K^* \gamma$. 

\section{Nonrelativistic $\bar Q Q$ Systems }

The top quark is the heaviest quark that has been observed. Over the last couple of years there has been important progress in our ability to predict the behavior of the $e^+ e^- \rightarrow \bar t t $ cross section near threshold. Several scales are important in this problem and that is what makes it difficult. If we neglect the weak interactions then toponium states would exist and the lowest lying states would be characterized by a relative velocity for the heavy quarks of magnitude $v$. The relevant physical scales are: $m_t$, $m_t v$ and $m_t v^2$. For $v=0.15$ these scales are $175~{\rm GeV}$, $26~{\rm GeV}$ and $3.9~{\rm GeV}$. With such a large difference between scales it is important to sum logarithms of the ratios of these scales. This is the only way one knows where to put the argument of the strong coupling and clearly $\alpha_s$ varies dramatically between these scales.
In the threshold region the cross section to $\bar t t$ (divided by the cross section to $\mu^+ \mu^-$) has the form
\begin{equation}
\nonumber
 R =
 v\sum\limits_{k,i} \left(\frac{\alpha_s}{v}\right)^k
 \left(\alpha_s\ln v \right)^i C
\end{equation}
where
\begin{eqnarray}
C=\left\{
 \begin{array}{ll}
 1 & \mbox{(LL)} \, \\
 \alpha_s, v & \mbox{(NLL)} \\
 \alpha_s^2, \alpha_s v, v^2 & \mbox{(NNLL)}
 \end{array} \right. 
 \label{RNNLLorders}
\end{eqnarray} 
The free cross section is of order $v$. The summation of Coulomb gluons is a power series in $\alpha_s/v$ and the renormalization group summation is in powers of $\alpha_s {\rm ln}v$. The logarithms are summed by going over to an effective field theory (NRQCD) with soft and ultrasoft gluons and using different subtraction points\cite{lms}, $\mu_s=m_t \nu$ and $\mu_{US}=m_t \nu^2$ for their interactions. One can think of $\nu$ as a subtraction point velocity.
\begin{figure}
\epsfxsize19pc
\epsffile[40 400 600 700]{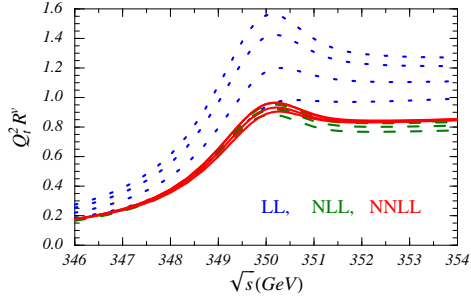}
\caption{Prediction for $R$ using renormalization group summation of logarithms of o $v$. At each order the plots are for
$\nu=0.1,0.125,0.2,0.4$}
\label{NLL}
\end{figure}

The prediction for $R$ using this formalism\cite{hmst} is shown in Figure (\ref{NLL}). At each order plots are shown for the subtraction velocity $\nu=0.1,~ 0.125,~ 0.2,~ 0.4$. The dotted lines have the leading logarithms summed (LL), the dashed lines have the next to leading logarithms summed (NLL) and finally the solid lines are at NNLL level. The plots are made from predictions where the top mass is eliminated in favor of half the $^1S_0$ toponium mass which is taken to be $175~{\rm GeV}$. The other parameters used are: $\alpha_s(M_Z)=0.118$ and the top width $\Gamma_t=1.43~{\rm GeV}$. It appears from the apparent insensitivity to the choice of $\nu$ that the cross section can be predicted to the $3\%$ level in the threshold region. If this is the case then it may be possible to determine the top mass at the $200~{\rm MeV}$ level, the top width to $5\%$ and the higgs $\bar t t$ coupling to $20 \%$ (for $m_H=115~{\rm GeV}$). The top mass comes mostly from the location of the peak, the top width mostly from the shape and the Higgs $t \bar t$ coupling mostly from the normalization. 

The prediction in Figure (\ref{NLL}) is a dramatic improvement over previous fixed order predictions\cite{h} where the logarithms are not summed in any systematic fashion. These had a fairly stable location for the peak but otherwise did not seem to be converging as one went from NLO to NNLO. 

There are some NNLL effects, which are thought to be small, that are not included in Figure (\ref{NLL}). Among these is a proper treatment of the top width at this order.

\section{Concluding Remarks }

There was no concluding "transparency" for my talk at Rome but I had meant to say a few words to put things into perspective. However, a tall, deeply tanned, well dressed Italian was hurrying me up and I decided it was in my best interest to forgo any conclusions. So I will take the opportunity here to write a few words. This has not been a classic review talk. No attempt was made to cover all the exciting developments that have occured over the last two years in this field. I have focused on a few areas where very important theoretical progress has been made and which in the future there will be dramatic experimental progress. 

With $\sin2 \beta$ recently measured at fairly high precision the next main target of the $B$ factories is likely to be improving our knowledge of $|V_{ub}|$. Remember the principal goal of the physics program of the $B$ factories is to over constrain the unitarity triangle, providing a precision test of the standard model in the flavor sector. It doesn't really matter in this program whether you measure a CP violating quantity or not. The length of a side is as good as an angle. What is really important is that precise measurements can be made with a clean theoretical interpretation.

It is possible we are on the verge of a systematic understanding of a wide class of exclusive nonleptonic $B$ decay amplitudes to light states. Given that we have failed miserably in our attempts to understand $D$ nonleptonic decays it would be particularly satisfying if the larger mass of the $b$-quark makes this possible in the $B$ system. Will the new ideas succeed and become a well justified quantitative tool? Its hard to say at this point, however, more experimental and theoretical progress will answer this question over the next few years.

An important part of the future of particle physics may be a very high energy linear $e^+e^-$ collider. Of course we hope that at such a machine we will be studying the properties of squarks, sleptons, Winos, {\it etc.} But its still nice to see the dramatic improvement that occured recently in our ability to predict the $\bar t t$ production cross section near threshold and there is some very interesting standard model physics to be done in this energy range. 

\section{Acknowledgments}
This work was supported in part by the Department of Energy under grant DE-FG03-92-ER-40701. I thank Z. Ligeti for some useful suggestions.


\begin{thebibliography}{99}
 
\bibitem{bll1} C. Bauer, Z. Ligeti and M. Luke, Phys. Lett. {\bf B479}, 395 (2000).

\bibitem{bn} M. Neubert and T. Becher, hep-ph/0105217 (2001) unpublished.

\bibitem{cleo1} S. Chen et. al. (CLEO Collaboration) hep-ex/0108032 (2001) unpublished; D. Cronin-Hennessy et. al. (CLEO Collaboration) hep-ex/0108033 (2001) unpublished.

\bibitem{bll2} C. Bauer, Z. Ligeti and M. Luke, hep-ph/010704 (2001) unpublished.

\bibitem{voloshin} M. Voloshin, Phys. Lett. {\bf B515}, 74 (2001).

\bibitem{llr} A. Leibovich, I. Low and I. Rothstein, Phys. Rev. {\bf D61}, 053006 (2000); A. Leibovich, I. Low and I. Rothstein, Phys. Lett. {\bf B513}, 83 (2001).

\bibitem{n} M. Neubert, Phys. Lett. {\bf B513}, 88 (2001).

\bibitem{opal} G. Abbiendi et. al. (OPAL Collaboration) Eur. Phys. J. {\bf C21}, 39 (2001).

\bibitem{bps} C. Bauer and D. Pirjol, I. Stewart, hep-ph/0107002 (2001) unpublished.

\bibitem{bbns} M. Beneke, G. Buchalla, M. Neubert and C. Sachadrajda, Nucl. Phys. {\bf B591}, 313 (2000).

\bibitem{dh} M. Diehl and G. Hiller, JHEP {\bf 0106}, 067 (2001).

\bibitem{llw} Z. Ligeti, M. Luke and M. Wise, Phys. Lett. {\bf B507}, 142 (2001).

\bibitem{cleo2} Alexander et. al. (CLEO Collaboration) Phys. Rev. {\bf D61}, 072003 (2000). 

\bibitem{cleo3} K. Edwards et. al. (CLEO Collaboration) hep-ex/0105071 (2001) unpublished.

\bibitem{cleo4} T. Coan, et. al. (CLEO Collaboration) hep-ex/0110055 (2001) unpublished.

\bibitem{belle} K. Abe et. al. (BELLE Collaboration) hep-ex/0109021 (2001) unpublished.

\bibitem{kls} Y. Keum and  H. Li, Phys. Rev. {\bf D63}, 074006 (2001); Y. Keum, H. Li and A. Sanda Phys. Rev. {\bf D63}, 054008 (2001).

\bibitem{b} S. Brodsky, hep-ph/0104153 (2001) unpublished.

\bibitem{cfmps} M. Ciuchini, E. Franco, G. Martinelli, M. Pierini and L. Silverstrini, Phys. Lett. {\bf B515}, 33 (2001); M. Ciuchini, E. Franco, G. Martinelli, M. Pierini and L. Silverstrini, hep-ph/0110022 (2001) unpublished.

\bibitem{bfs} M. Beneke, T. Feldmann and D. Seidel, Nucl. Phys. {\bf B612}, 25 (2001).

\bibitem{bb} S. Bosch and G. Buchalla, hep-ph/0106081 (2001) unpublished.

\bibitem{lms} M. Luke, A. Manohar and I. Rothstein, Phys. Rev. {\bf D61}, 074025 (2000).

\bibitem{hmst} A. Hoang, A. Manohar, I. Stewart and T. Teubner, Phys. Rev. Lett. {\bf 86}, 1951 (2001).

\bibitem{h} A. Hoang, et. al., Eur. Phys. J. {\bf C3}, 1 (2000).


\end{thebibliography}
\end{document}